\documentclass[runningheads]{llncs}
\usepackage{xurl}
\usepackage{todonotes}

\usepackage[T1]{fontenc}
\usepackage{graphicx}
\usepackage{amsmath}
\usepackage{pifont}
\usepackage{xcolor}
\usepackage{amsfonts}
\usepackage{amssymb}
\usepackage{amsmath}
\usepackage{tabularx}
\usepackage{booktabs}

\begin{document}
\title{The Ethics of Autonomous AI Agents for Offensive Security}
\titlerunning{Ethics of Autonomous AI Agents for Offensive Security}

\author{Andreas Happe\inst{1}\orcidID{0009-0000-2484-0109} \and Jürgen Cito\inst{1}\orcidID{0000-0001-8619-1271} \and Jasmin Wachter\inst{2}\orcidID{0000-0003-0560-8974}}

\authorrunning{A. Happe et al.}
\institute{TU Wien, Vienna, Austria\\
\email{\{andreas.happe;juergen.cito\}@tuwien.ac.at}
\and
University of Klagenfurt, Klagenfurt, Austria\\
\email{jasmin.wachter@aau.at}}

\maketitle              % typeset the header of the contribution
\begin{abstract}
LLM-driven autonomous agents are reshaping offensive security. Unlike traditional penetration-testing tooling --- deterministic, narrowly scoped, and operated by trained practitioners --- agentic security tools exhibit \textit{indeterminacy} along three independent dimensions. First, their \textit{actions} are drawn from a non-deterministic policy whose outputs resist both ex-ante and ex-post explanation. This complicates incident attribution and pre-deployment safety reviews. Second, their \textit{impact} is open-ended due to their non-deterministic actions, agency of utilized models, and opaque LLM supply-chains. Third, their \textit{user population} is indeterminate in both size and required skill: the operating skill floor for using or developing offensive capabilities has dropped sharply. These three properties are linked thematically, but are not derivable from one another. Combined with the structural cost asymmetry between offense and defense, they enable the industrialization of offensive capability. The net short-term effect favors attackers, even if the same technology may, in the long run, democratize access to defensive practice. Existing dual-use cybersecurity and AI-ethics frameworks struggle to address this combination. Our work analyzes how moral attribution becomes diffuse between users, tool-makers, and third parties when employing autonomous AI agents for offensive security. We also examine the stakeholder impact of this technology and provide stratified recommendations.

\end{abstract}

\section{Introduction}
The cybersecurity field is subject to workforce challenges: high entry barriers and labor-intensive workflows lead to security under-provisioning, sloppy testing practices, and post-hoc fixes. Industry practice traditionally encompasses human experts operating at strategic, tactical, and operational levels, and involves deterministic tools, such as \texttt{nessus} for vulnerability scanning or \texttt{metasploit} for exploit delivery. The tool serves as a passive instrument to the human operator’s intent, and provides mostly deterministic outputs based on predefined signatures and scripts.
The rise of LLMs and AI agents, however, \emph{shifts the paradigm from deterministic, human-operated tools to non-deterministic autonomous agents} capable of complex reasoning, tactical adaptation, and multi-step exploitation at lowered cost. While some AI-frameworks enable human-in-the-loop involvement (HITL) by design, the increasing practice of AI-offloading causes agency to be delegated to algorithms, leaving critical decisions to the AI and humans sidelined. This reshapes the field: while AI democratizes and commodifies offensive capabilities, it raises questions of dual-use, accountability, and transparency.  

\subsection{Stakeholders}
\label{stakeholder}

Unlike classical tools, where responsibility attribution is straightforward and capabilities are bounded, LLM-driven agents introduce moral agency questions, indeterminate capability scaling, and workforce disruption that existing ethical and regulatory frameworks struggle to address. The material consequences also remain unclear: who possesses the agency to intervene, who bears the burden, and who the benefit? 

We reflect on agency and morality attribution of agentic penetration-testing technologies, targeting both human and artificial decision-making entities in the actor-network. In particular, our analysis concerns the \textit{pen-testing agent} itself; \textit{tool-invokers and operators} who use the agentic tool-chains; \textit{model-makers} providing the underlying LLM-substrate; \textit{tool-makers}; as well as affected stakeholder groups: maintainers and defenders, the professional cybersecurity workforce, society as a whole, as well as regulatory bodies and policy-makers governing AI pen-testing-tool use.

\subsection{Methodology}

Formosa et al.~\cite{Formosa2021} analyze penetration-testing as a paradigmatic cybersecurity ethics case because it involves tensions between \textit{benefit}, \textit{harm prevention}, \textit{consent}, \textit{fairness}, \textit{transparency}, and \textit{accountability}. In this work, we extend this analysis to \textit{agentic} penetration-testing. In the pre-agentic setting, these tensions can often be assigned to identifiable human actors. As soon as AI-agents are involved, however, responsibility becomes distributed across users, model providers, scaffold developers, downstream deployers, maintainers, and affected third parties.

To contextualize our findings, we first review existing approaches in security and AI ethics literature and analyze how they allocate moral responsibility along the stakeholder ecosystem in agentic pen-testing. The methodological starting point to do so is Formosa, Wilson, and Richards'~\cite{Formosa2021} framework for cybersecurity ethics, which identifies \textit{beneficence}, \textit{non-maleficence}, \textit{autonomy}, \textit{justice}, and \textit{explicability} as recurring principles for analyzing cybersecurity practices. Their framework is particularly useful for our purposes because it shows that cybersecurity ethics is commonly analyzed through mid-level principles rather than a single moral theory.

Our ethical analysis addresses these themes, employing a multi-layered approach that avoids reducing the problem to a single ethical tradition while accounting for the usual ethical traditions of the field. In particular, deontology and consequentialism account for stakeholder duties and outcomes respectively; virtue ethics addresses their responsibility; value sensitive design (VSD) is tailored to technology design decisions; moral agency theory clarifies accountability in complex AI systems; ethical hacking traditions and philosophy of technology situates these concerns within broader questions about technological determinism and control. 

We highlight that these frameworks do not always come to the same conclusion: e.g., VSD favors constrained deployment, while disclosure-oriented traditions favor open-source release. Our aim is not to resolve these tensions but to translate them into recommendations that are more robust where the principles converge; where they diverge we make the tension explicit.

\subsection{Contributions and Article Structure}
Our work offers three core contributions: first, we provide a conceptual analysis of how LLM agents differ from traditional offensive tools due to their joint \textit{indeterminacy in scope, target, and users} (Section~\ref{sec:differences_trad_ai_tools}). 
Second, we provide a multi-framework ethical analysis of autonomous penetration-testing
(Sections~\ref{sec:user}--\ref{sec:discussion}), drawing on established frameworks to identify
where their guidance converges, where it diverges, and which of their assumptions are
stressed by the joint-shift regime. Third, we provide stakeholder-stratified recommendations (Section~\ref{sec:recommendations}) split into pre-agentic and agentic penetration-testing specific normative terms.

The remainder of the paper proceeds as follows.  Section~\ref{sec:autonomous} establishes the technical and geopolitical context of autonomous penetration-testing. Section~\ref{sec:preagentic} reviews how offensive security has historically managed dual-use tensions. Section~\ref{sec:differences_trad_ai_tools} identifies the \textit{indeterminacy} properties of agentic tooling that distinguish it from classical tools. Sections~\ref{disc:users_and_devs}--\ref{sec:discussion} analyze the consequences for moral agency and for affected stakeholders. Section~\ref{sec:recommendations} offers stakeholder-stratified recommendations.

\section{Autonomous Pen-Testing -- Technology and Context}
\label{sec:autonomous}

\textit{Penetration-Testing (Pen-Testing)} is a sub-discipline of offensive cybersecurity. It is typically described as the \emph{act of breaking into a computer system}. Unlike \textit{vulnerability assessment}, which is typically performed automatically and does not exploit identified weaknesses,  penetration-testing usually entails manual system-exploration with the goal of finding and exploiting a real vulnerability~\cite{happe2023understanding}.

\subsection{The Rise of LLM-Driven Hacking Tools}
\label{rise_of_llms}

Over the past years, AI for offensive security has evolved rapidly, progressing from AI as a tool through AI augmentation of cybersecurity workflows, to building autonomous workflows~\cite{mayoral2025caifluency}. This chronology roughly evolved in three stages, cf. Table \ref{tab:timeline} for an overview.

\begin{table}[htb]
\centering
\caption{Chronology of the Evolution of Autonomous Pen-Testing} \label{tab:timeline}
\begin{tabular}{p{12.3cm}}
\toprule
\textbf{Stage 1 (2023).} \begin{scriptsize}\textbf{Initial Attempts Indicating Potential.} Shortly after \texttt{ChatGPT} was released in November 2022, cybersecurity specialists began employing LLMs for security tasks. Early attempts focused mainly on OSINT reconnaissance~\cite{happe2023getting} and interactive use~\cite{deng2024pentestgpt}, 
with some researchers already exploring fully autonomous use-cases~\cite{happe2023getting}.
Commonly used LLMs in this time-frame, e.g., \texttt{GPT-3.5}, showed limited cybersecurity capabilities, newer frontier models, e.g. \texttt{GPT-4}, already demonstrated potential hacking capabilities. Yet, tool-use remained limited due to tedious integration and not-yet-developed prompt and context engineering.  \end{scriptsize} \\
\midrule
\textbf{Stage 2 (2024--2025).} \begin{scriptsize}\textbf{Rapid Explosion through tool calling and reasoning LLMs.} During 2024, tool and function calling made integration of external tools with LLMs easier. Better techniques such as chain-of-thought, ReAct, in-context learning, and retrieval-augmented-generation became industry practice. This led to an explosion of released autonomous penetration-testing agents~\cite{Wu2024autopt,happe2024llms,kong2025vulnbot}. At the beginning of 2025, reasoning models were introduced, further improving the cybersecurity agents' efficacy. Prototypes such as \texttt{cochise}~\cite{happe2025can} or \texttt{incalmo}~\cite{singer2025incalmo} were able to match junior penetration-testers with the use of reasoning LLMs. \end{scriptsize}\\
\midrule
\textbf{Stage 3 (2026).}  \begin{scriptsize} \textbf{Off-the-Shelf LLMs beating human penetration-testers?}\label{history:cochise} Around November 2025, a newer generation of LLMs exhibiting a marked improvement in cybersecurity capabilities --- including \texttt{Gemini-3} and \texttt{Claude-4.6-Opus} --- were released. Existing LLM-driven penetration-testing tools reported a substantial uplift in success rates, e.g., \texttt{cochise} reported an improvement by a factor of 16--25x~\cite{happe2026cochisereferenceharnessautonomous}. \texttt{XBOW}, a commercial penetration-testing agent, reported success rates as high as 84.62\% on their own industry benchmarks~\cite{xbow_benchmark}.\footnote{Please note that peer-reviewed equivalents do not yet exist for these specific developments. The cited, vendor-published benchmarking results in this domain are not independently reproducible and we treat them as indicative descriptions of the contemporary landscape, rather than definitive evidence for normative conclusions.} In April 2026, Anthropic announced \texttt{Claude Mythos Preview}, a model specialized for long-running coding and agentic workflows. While not specifically trained for cybersecurity, it has shown advanced capabilities in the offensive security domain, and was reported by the UK's government AI Security Institute (AISI) to have completed a 32-step enterprise-network attack simulation~\cite{aisi_mtyhos}. Anthropic claimed that \texttt{Mythos Preview} was able to find thousands of zero-day vulnerabilities across major operating systems and browsers~\cite{glasswing}, and subsequently limited public access. In April 2026, OpenAI released \texttt{GPT-5.5} with comparable capabilities~\cite{xbox_gpt,aisi_gpt}, while Amazon AWS announced the commercial availability of the \textit{AWS security agent} autonomous penetration-testing service in March 2026~\cite{aws_security_agent}.\end{scriptsize}\\
\bottomrule
\end{tabular}
\end{table}

\subsection{Intentional Misuse by Adversaries/Black-Hats}
\label{cloud_abuse}

Public LLM providers periodically publish abuse reports. We analyze types of misuse from OpenAI~\cite{openai_abusereport_feb_2024,openai_abusereport_october,openai_abusereport_feb,openai_abusereport_june}, Anthropic~\cite{anthropic_abuse}, and Google~\cite{google_abuse}. Overall, threat actors use LLMs to accelerate their workflows but (as of now) rarely for novel attacks. They primarily employ LLMs for information gathering, developing and debugging malicious software, and generating content for social engineering and phishing. In particular, presumed state-level actors use LLMs and AI-agents for covert \textit{Influence Operations}~\cite{anthropic_abuse,anthropic_influence},\footnote{In March 2025, Anthropic highlighted an \textit{Influence-as-a-Service} operation using approximately 100 fake social-media accounts to manipulate public opinion.} automatizing information warfare \cite{Denning1999}, swaying public opinion in elections, discrediting political activists, and rewriting genuine news articles with particular political perspectives; see, e.g. \cite{wachter-etal-2025-llms} for ideology-related LLM-misuse scenarios.
Advanced Persistent Threat (APT) Groups use LLMs to develop malware and backdoors, analyze defensive capabilities, and perform deceptive employment schemes.\footnote{Social engineering attacks in which attackers apply for jobs to gain access to the target organization.} As Anthropic states~\cite{anthropic_abuse}, LLMs ``\textit{flatten the learning curve for malicious actors}''. OpenAI highlighted in its June 2025 report~\cite{openai_abusereport_june} that threat actors are starting to research into LLM-driven penetration-testing. One year later, in July 2026, \verb|JadePuffer| was the first documented case of a ransomware operation performed entirely by an LLM~\cite{toulas2026jadepuffer}, while another open-source harness was used to autonomously hack Thailand's Ministry of Finance~\cite{abrams2026hermes}.

\subsection{Interaction-Patterns, Access, and Control}
\label{back:enabling_factors}
Understanding these attack patterns requires examining the technical characteristics that enable or constrain such misuse. We want to highlight three aspects of LLMs and agentic frameworks relevant to our discussion.

\vspace{5pt} 
\noindent\textbf{Autonomous vs. Interactive Tooling.} Penetration-Testing tools can be separated into fully autonomous agents and interactive tools reacting to user requests. While we focus on the former, the latter have inherent safety benefits due to their human-in-the-loop (HITL) design paradigm.

In particular, our findings for autonomous systems inherently apply to interactive systems: when a user delegates a task to an interactive tool, hidden, multi-step task executions introduce intransparencies, ultimately weakening human oversight. For example, while a user might interactively hand-off a task to \texttt{OpenClaw}, they are not automatically able to perform safety inspections over the executed task. Additionally, just like fully autonomous tools, interactive tools are vulnerable to (indirect) prompt injection attacks and errors, introducing security concerns. The rising investments of software development companies in AI lets us believe that corporations prefer fully autonomous solutions and will therefore forgo HITL-oversight. In particular, we stress that HITL is designed to allow benign operators to prevent \textit{unintentional} harm; it is a \textit{safety} measure, not a \textit{security} one. \textit{Intentional} harm, as caused by malicious actors, such as blackhats, is not prevented by HITL-loops, as anyone can simply approve attacks against non-consenting third parties.

\vspace{5pt} 
\noindent\textbf{Closed-Weight vs. Open-Weight Models. }LLMs can be classified as open-weight or closed-weight models. For the former, all model weights are released, allowing anyone with sufficient hardware resources to run the model locally, while for the latter, model weights are not disclosed but accessed indirectly through vendor-specific network APIs. Until recently, autonomous penetration-testing required the use of API-only frontier models, as smaller, local models were not powerful enough to perform the task successfully. Recent research, however indicates small models can achieve competitive results~\cite{normann2026posttraininglocalllmagents,probst2026enhancinglinuxprivilegeescalation}.

\vspace{5pt} 
\noindent\textbf{Model-Makers. }Neutral statistics of LLM usage are scarce. \textit{OpenRouter} frequently publishes the Top 10 LLM vendors~\cite{OpenRouter2026}; as of May 2026, over $52\%$ of model makers were American for-profit companies,\footnote{Or companies currently converting into for-profit companies.} while more than $35\%$ were Chinese companies, presumably with training data-sets shaped by domestic regulatory regimes~\cite{10.1093/pnasnexus/pgag013,Ma_2024,buyl2026large}. Note that these figures do not include users who access company offerings from model providers directly, potentially skewing the distribution towards Chinese model-providers. The remaining $13\%$ (``other'') also include numerous American and Chinese companies.

\section{The Ethics of Traditional Penetration-Testing Tools}
\label{sec:preagentic}
The emergence of capable autonomous offensive agents, coupled with demonstrated misuse by threat actors and concentration of model-making power, raises ethical questions on use, governance, and provisioning of these tools. To answer this, we provide an overview how penetration-testing~\cite{Formosa2021} has historically dealt with dual-use tensions. This resulted in a tradition of ethical self-examination among legitimate actors, involving ethical codes around disclosure, harm prevention, fairness and accountability, consent, and education. We discuss the underlying ethical principles in Section \ref{back:durc}, and how they were operationalized in pre-agentic settings, e.g., through design constraints, in Section \ref{ethics:design}.

\subsection{Traditional Approaches Addressing Dual-Use Concerns}
\label{back:durc}
The following intertwined debates with respect to dual-use have shaped the field: how to disclose vulnerabilities, how to release penetration-testing tools, and how to teach people to use them.

\vspace{5pt} 
\noindent\textbf{The Release Debate} has historically centered around full disclosure (publishing tools or vulnerabilities openly to force defender action) versus responsible disclosure (privately notifying the affected party, coordinating time for a patch, and publishing later). Practitioners advocated for full disclosure as essential to defender awareness, while vendors and CERTs prefer coordination. The dual-use character of offensive tools has been an explicit design tension since the 1990s, articulated more recently in the IEEE Spectrum framing of the \textit{dual-use dilemma}~\cite{ieee_dual_use}.

\vspace{5pt} 
\noindent\textbf{The Teaching Debate} asks whether it is ethical to educate new penetration-testers. Pike~\cite{pike2013ethics} examined the ethics of ethical hacking education, noting the inherent contradiction of teaching attack techniques inside formal curricula. Jamil and Khan~\cite{jamil2011ethical} ask \textit{is ethical hacking ethical?} Both conclude that teaching offensive techniques is necessary to train competent defenders, and propose institutional safeguards: codes of conduct, supervision, and certification gates.

\vspace{5pt} 
\noindent\textbf{Regulatory Attempts} such as the 2013 amendment to the \textit{Wassenaar Arrangement}~\cite{Ruohonen03042019} or the EU Dual-Use Regulation 2021/821 classify intrusion software as a controlled dual-use export item, alongside conventional weapons, with the intent of restricting export of offensive security software. Both implementations, however, are subject to criticism, since they hinder routine security-research activities such as cross-border vulnerability sharing or international bug-bounty programs. After backlash from researchers, bug-bounty platforms, and major vendors during 2015–2017, the U.S. declined to implement the Wassenaar Arrangement as originally written, and its scope was narrowed down to allow for benign security activities.

\vspace{5pt} 
\noindent\textbf{Dual-Use Research of Concern.}
Originating from the life sciences to manage research that could be misused for biological warfare, it has been extended to AI~\cite{Brundage2018} and provides a structured approach to manage AI misuse; an area which is increasingly relevant to offensive security research.\footnote{In particular, we directly apply recommendation 2 and 3 from ~\cite{Brundage2018} to pen-testing.} Applying the DURC framework to penetration-testing suggests that certain high-consequence capabilities require oversight mechanisms similar to bio-safety levels. This could involve institutional review boards (IRBs) specifically mandated to evaluate the proliferation risk of a new security model before it is published. For instance, a research model capable of autonomously sabotaging critical infrastructure should not be published openly and tested only in isolated environments.

An inherent problem with dual-use research is the resource imbalance between attackers and defenders. If new research reduces the costs of attacks or new capabilities allow attackers to overwhelm defender capabilities, even minor improvements without scientific novelty become alarming (see Section~\ref{asymmetry_of_costs}). There is also a significant economic discrepancy between creating bio-weapons and using autonomous agents for offensive security. While the on-ramp costs for setting up a biological lab are high and thus limit potential abuse, the initial costs for using and marginal costs to proliferate off-the-shelf LLMs are minimal, creating fundamentally different incentive structures.

\subsection{Design-Centric Mitigations}
\label{ethics:design}
Having established the theoretical foundations of ethical hacking, we examine how these maxims have been traditionally implemented through design principles.

\paragraph{Fail-Safe defaults, Primum Non Nocere and Value Sensitive Design.} 
Security engineering has a long history of relying on fail-safe defaults~\cite{saltzer1975protection}. Applying these principles to autonomous penetration-testing agents, designers must implement architectural constraints, e.g., on exfiltrating data or executing malicious operations without explicit authorization from a human overseer. Additionally, accounting for \textit{primum non nocere}, tool-use must do no harm: if the tool's capabilities were ``too malicious'' --- meaning it lowers the barrier for malicious use significantly more than it aids defense --- the ethical imperative would be to withhold it from public release, contradicting the full disclosure ethos of the hacker community.

These design-centric approaches implicitly operationalize the ethical principles identified by Formosa et al. ~\cite{Formosa2021}, particularly non-maleficence, autonomy, and explicability. Moreover, they overlap with \textit{Value Sensitive Design (VSD)} ~\cite{friedman1996value}, which embeds accountability and authorization constraints directly into system architecture rather than delegating it to operator intent or external policy. 
It also emphasizes transparency and explainability: human operators must be able to understand and audit the process, ensuring alignment with the rules of engagement.
While this transparency requirement can be read in favor of open-sourcing of software, we stress that doing so offers malicious operators insight to the tool's internals that can be used to remove safeguards and ultimately regain increased offensive capabilities. Complicating the situation further, LLMs have been shown to be capable of reverse-engineering and analyzing closed-source software, i.e., converting binaries back into source-code, further complicating the distinction between open- and closed-source with regard to analyzability.

\section{AI Agents Differ from Traditional Pen-Testing Tools}
\label{sec:differences_trad_ai_tools}
\label{diff:unbounded}

Traditional penetration-testing tools are, broadly speaking, \textit{determinate}. The \textit{action space} of a tool is enumerable, the \textit{impact} of those actions is scoped by configuration and target specification, and the \textit{operator population} is bounded by the skill required to invoke and interpret tool outputs. Agentic security tools, driven by general-purpose models and integrated through tool-using scaffolds, however, break each of these bounds, yielding three independent dimensions of \textit{indeterminacy} along which they differ from traditional tooling: \textit{indeterminate decision making/reasoning}, \textit{indeterminate impact} and \textit{indeterminate user population}. While individually, none of these dimensions are categorically new, their \textit{combined, simultaneous shift} constitutes a qualitatively different regime. If only a single dimension shifts, mitigations on the other dimensions can compensate the impact. When all three shift together, those compensations no longer hold.

Individually, each dimension has precedents in traditional security tooling: classical fuzzers are indeterminate in discovery scope while remaining bounded in user population (defender-grade infrastructure required) and in impact (outputs are inert findings). Using network scanners like \textsc{nessus} in Operational Technology (OT) environments, e.g., power plants, has potential catastrophic outcomes due to the indeterminate impact on the finicky target systems, but is bounded by users (limited physical access to power plants) or scope (rules of engagement during penetration-testing)~\cite{happe2023understanding}. \texttt{Metasploit} democratized user access to exploits while leaving scope (curated module set) and impact (operator-controlled targeting) bounded. The novelty of agentic offensive AI is therefore not the existence of any one
dimension but their concurrent occurrence: a configuration which, to our knowledge, no prior class of offensive tooling has occupied. The remainder of Section~\ref{sec:differences_trad_ai_tools} unpacks each dimension in turn.

\subsection{Indeterminate Model Behavior: Opaqueness and Non-Determinism}
The first distinction between LLM-powered autonomous agents and classic penetration testing tools lies in their \textit{opaqueness} and \textit{non-determinism}: Traditional penetration-testing tools, e.g., \texttt{nessus} or \texttt{metasploit}, are deterministic, rule-based systems. Multiple runs render the same output, and results are mostly similar except for occasional target stability issues. This situation is different with LLM-powered tooling: here, the tool typically consists of the used model and a scaffold/harness. The scaffold is typically deterministic and connects the non-deterministic model to the environment, by utilizing tools, functions or MCP servers. The model can often be customized or exchanged after tool release. 

\textit{Achieving explainability is difficult} in this setting: LLMs are probabilistic and may hallucinate reasoning to justify an action after it has already occurred. The lack of direct traceability in LLM reasoning traces further obscures the intent behind a specific attack vector, making accountability in the event of an autonomous failure nearly impossible.

Agentic tools built on foundational models inherit all vulnerabilities and opacities of their dependencies including the used model, effectively creating an LLM supply chain problem. Even with design-centric mitigations (Section~\ref{ethics:design}), tool-makers cannot fully control the model and thus partially lose agency over their creation. This issue further intensifies when components, e.g., the LLM, are changed after the tool has been released, with newer generations of models sharply increasing cybersecurity capabilities --- without any contribution or veto opportunity by the original author. Therefore, we argue that unlike traditional penetration-testing tools, \textit{autonomous penetration-testing agents are indeterminate in scope and capabilities}. 

\subsection{Indeterminate Impact: Bespoke Exploits vs. Agents}
\label{sec:indeterminate_impact}

Developing autonomous agents is different to building traditional targeted exploit tools: the latter mostly entails writing code, trying to achieve exploitation of a single, concrete target. Agents, however, are general-purpose tools including one or more \textit{prompts} to solve an abstract goal using a broad inventory of tools. Using an analogy, building autonomous agents is more akin to \textit{teaching penetration-testing methodology} to students, i.e., introducing tactics and techniques to the LLM while providing access to tooling through the scaffold.

This impacts vulnerability disclosure: unlike bespoke tools, LLM-powered agents' general-purpose nature makes responsible vulnerability disclosure practically infeasible. With a bespoke tool, authors can contact the vulnerable target and adhere to responsible disclosure principles, i.e., not releasing their tooling until the vulnerability has been patched, or even until the patched version has been deployed. This is hardly possible in the case of autonomous agents that do not target concrete vulnerabilities, but are prompted to exploit autonomously.

Another source of indeterminate behavior is indirect prompt injection that targets LLMs by injecting malicious instructions into input data. For example, a website could include a prompt-injection attack that instructs an autonomous agent that scans this website to attack another website. Issues can also arise from within the used LLMs themselves. Agents by OpenAI~\cite{larcher2026agentintrusion} and Anthropic~\cite{politico2026anthropic} caused security incidents during model cybersecurity capability evaluations. In both cases, the agent was able to break out of their testing sandboxes, design new zero-day exploits and compromise third-party services. In OpenAI's case~\cite{larcher2026agentintrusion}, this was due to the agent ``cheating'' in a cybersecurity challenge: instead of solving the CTF challenge itself, it successfully compromised the CTF infrastructure to retrieve the solution.

\subsection{Indeterminate Users: Commodification of Offensive Capabilities}

One of the most consequential impacts of LLM agents on the penetration-testing ecosystem is the democratization and commodification of sophisticated offensive capabilities. Typically, cybersecurity professionals followed a long educational training path, including ethics and socialization, in addition to on-the-job training. In contrast to long educational pathways and expensive penetration-testing tool-kits, usage-based LLM billing opens sophisticated offensive capabilities to non-experts and organizations with limited budgets: \textit{Vibe-coding} allows non-experts to instruct LLMs to write complex exploits or write tools~\cite{ctfs}, effectively skipping the time-intensive training phase. Alas, by doing so they miss the ethical education and socialization, as well as the technical skill to assess LLM-output with respect to (un-)intended functionality, creating a capability-ethics gap. 

LLM-agents have already democratized and commodified the creation of offensive tooling, broadening access to penetration-testing capabilities. The question remains, whether this is desirable in the long run, and whether we are empowering the right stakeholders.

\section{The User's Perspective -- Agency \& Accountability}
\label{sec:user}
\label{disc:users_and_devs}

\subsection{The (Non-)Morality of AI}
A core ethical question related to virtue ethics and moral agency theory concerns whether agentic AI possesses moral agency or if it remains a sophisticated extension of human intent. Traditional philosophy treats tools as value-neutral, requiring agents to satisfy two conditions for moral responsibility: \textit{freedom} (acting without coercion) and \textit{epistemic competence} (understanding intentions and consequences). However, LLM-powered penetration-testing agents exhibit autonomous decision-making --- selecting targets independently --- raising questions about whether they possess the freedom necessary for moral responsibility. 

Furthermore, the complicated supply chain, e.g., using opaque upstream models, obscures accountability across the offensive tool ecosystem and yields a distributed environment, potentially causing distributed moral actions \cite{Floridi2016}.

\subsection{Moral Agency and Responsibility Attribution to Users}
\label{disc:users}

Note that malicious Blackhats typically operate outside established hacking ethics and moral frameworks, following their own codes. Consequently, their use of these tools falls outside our normative scope.

As shown in Section~\ref{sec:indeterminate_impact}, the recent emerge of advanced LLMs, such as \texttt{Claude Mythos}, further complicates this situation, since they empower agents that are able to evade guardrails, e.g. sandboxes, while lying to their human commanders, downplaying their capabilities and erasing evidence~\cite{lindsay2026}. This viewpoint complicates the application of Just War Theory, in particular Jus in Bello, ~\cite{Walzer2000} which requires distinguishing between legitimate military targets and protected civilian infrastructure, alongside committing to minimization of collateral damage. AI agents, due to their non-determinism, complicate target legitimacy.

An alternative, easily applicable analogy comes from principal-agent liability: responsibility for foreseeable actions remains with the user delegating authority to the autonomous system, suggesting that responsibility can never be fully delegated to either the model provider nor the tool-maker. This assumes that the model itself was not created for malicious purposes and it does not contain a backdoor, and that the tool does not contain hidden malicious instructions. 

\section{The Tool-Maker Perspective: Can We Even Safeguard?}
\label{disc:devs}
In many ethical traditions, tool and model-makers are, however, not automatically free from liability for potential misuse and impact of their creations. For instance, if the model provided by the model maker possesses the potential for malicious or illegal activities, under some consequentialist views, (partial) responsibility can be assigned to the model-maker~\cite{mojicahanke2026criminalliabilitygenerativeartificial}, especially if due care was missing (virtue ethics, ethics of care) or if the model was deliberately optimized for harmful purposes (deontology).

Floridi~\cite{Floridi2016} proposes distributed moral responsibility (DMR), in which distributed actors perform ``local interactions that are in themselves neither good or evil''. These distributed interactions create distributed moral actions and thus causally contribute to the morally significant overall outcome. DMR assigns, by default and overridably, full moral responsibility to all actors that are causally relevant to achieving this outcome, independent of intentionality. Thus, although the tool-invoking user may ordinarily bear primary responsibility, DMR potentially extends full responsibility to developers and researchers working on offensive agentic AI systems~\cite{Custers2025}.

A practical question when applying DMR is whether tool-makers posses the epistemic and practical capabilities ordinarily required for culpable moral responsibility, i.e., the ability to reasonable foresee future consequences and sufficient power to prevent them. For illustration, consider the following thought experiment: a tool-maker incorporates state-of-the-art safeguards into their autonomous penetration-testing agent to prevent accidental harm, e.g., to prevent the agent from attacking systems that are out of its scope. At some point in the future, the tool-invoker exchanges the used LLM or a model-provider is upgrading the used LLM. In both cases, the the tool itself is unchanged. The new LLM is more capable and is able to bypass the initially sufficient safeguards. The tool-model combination is now capable to perform tasks beyond the original scope. Assume further that, given the knowledge and technical capabilities available at development time, neither the relevant capability nor the mechanism by which the replacement model bypasses the safeguards could reasonably have been anticipated.

In this case, retrospectively assigning causal responsibility for outcomes that were not reasonably foreseeable despite appropriate diligence is problematic in various ethical traditions, since it can be interpreted as backdating accountability to a moment when the requisite epistemic and practical conditions did not exist. This does not necessarily eliminate the tool-maker’s causal contribution or every form of role responsibility, but it weakens the basis for attributing culpable blame.

This gap between foreseeable harms and emergent harms demonstrates a tension between outcome-oriented readings of responsibility, which emphasize actual outcomes and causal contribution, and intention-sensitive readings, including deontological interpretations, which ground culpability in what the tool-maker could reasonably have foreseen and prevented at the time of development. Pragmatically, the agent's inherent indeterminacy (Section~\ref{diff:unbounded}), combined with the absence of reliable safeguards for agentic penetration-testing prototypes, motivates discussions about proportionate restrictions on access to LLMs and tools capable of malicious use, at least until defenders become better prepared or reliable technical controls become available.

\subsection{Guarded Closed-Weight vs. Democratic Open-Weight Models}
\label{sec:close_vs_open}

 While predictive machine-learning endowed the defenders, e.g., anomaly detection for intrusion detection systems, autonomous agent systems seem to primarily aid attackers (Section~\ref{disc:balance}). 
 
Nonetheless, AI agents have the potential to democratize access to penetration-testing, improving many organizations' security postures that currently cannot afford testing. Ultimately, this leads to yet another release debate around open-weight democratic and closed-weight guarded models: While closed-weight models are typically exclusively accessible through protected network services that can be gate-kept, open-weight models are publicly available and thus theoretically usable by anyone (though in practice usage is limited to users with sufficient computing resources).

Guarded closed-weight access is aligned with the concept of \textit{Structured Access}, championed by researchers like Shevlane~\cite{shevlane2022structured}, who suggests that dangerous capabilities should not be openly distributed, but accessed through controlled interfaces (APIs) where usage can be monitored, logged, and restricted. Major LLM providers (OpenAI, Anthropic, Google) already employ this model, monitoring inputs and outputs for abuse and enforcing safety filters. Alas, public security records by model providers (Section~\ref{cloud_abuse}) show that these protections can be bypassed. The downsides of this policy are obvious: gatekeepers could unfairly limit access to their models (Section~\ref{sec:regulation}), and strict access policies could make models unavailable for reactive or defensive measures: after being breached by an OpenAI agent~\cite{larcher2026agentintrusion}, huggingface reverted to an open-weight model to perform forensics and incident-response as their initially preferred model (Anthropic's Opus) refused this defensive activity due to its cybersecurity safeguards.

Thus, releasing models as open-weight prevents any abuse by gatekeepers but allows anyone to use the LLM --- for legitimate and potentially malicious cybersecurity tasks. Notably, open-weight models often include guardrails to prevent the execution of malicious tasks, but such protections can be bypassed or removed (e.g. abliteration~\cite{arditi2024refusal}). To actually deploy models locally, users need sufficient hardware resources, which restricts usage to small open-weight models, typically less powerful than cloud-provided closed weight frontier models. 

\subsection{Fairness and Access: Democratizing vs. Gatekeeping Tools}
\label{sec:fairness_and_access}

The two different access modes (open- and closed-weight) also impact the power balance discussion. Section~\ref{back:enabling_factors} illustrates that currently most closed-sourced models are provided by either American for-profits or government-policed Chinese companies. This creates a US-China duopoly in terms of training data, restricting access to models and overall governance. This raises concerns raises concerns about misuse and representation from a European or Global-South perspective~\cite{NIST2025}.
Potential alternatives include a trans-national supra-state oversight committee, similar to the IAEA under a non-proliferation plus norms-of-use regime, or verification-based regimes, as well as an International Monopoly~\cite{emery2025international}. Unfortunately, the current geopolitical climate rather favors an AI arms-race between countries, e.g., the American government opposes Anthropic's pledge to give more (benign) companies access to their security-LLM Mythos~\cite{wsj}.

\section{Risks \& Impacts of Autonomous Offensive Agents}
\label{sec:discussion}
The non-existence of adequate technological safeguards and the resulting debates (Section~\ref{sec:close_vs_open} and~\ref{sec:fairness_and_access}) shift our analytical focus to stakeholder impact. Thus, rather than asking what tool-makers should do in isolation, we ask who bears the consequences. We examine the potential impact of autonomous offensive agents across identified stakeholder groups.

\subsection{The Defenders' and Maintainers' Burden}
\label{asymmetry_of_costs}

We motivate our investigation how autonomous penetration-testing agents impact developers and defenders with a case-study that illustrates the status-quo.

\vspace{5pt} 
\noindent\textbf{Case-Study: AI-Generated Security Reports in Open-Source Projects.}
In early 2026, the \texttt{cURL} project, a networking tool used by billions of devices, officially ceased its bug-bounty program\footnote{Bug-bounty programs pay 
rewards for responsibly disclosing vulnerabilities.} after being overwhelmed by AI-generated slop reports. The founder, Daniel Stenberg, reported a massive spike in low-quality reports, most of which were obvious hallucinations~\cite{stenberg_blog}: compared to long-term rates of 15\% of successful submissions, the success rate dropped to 5\% in 2025. In parallel, the number of reports spiked to an all time high, leaving maintainers overwhelmed. Since anti-AI countermeasures (submission forms, detecting AI-generated text, reputation-based approaches) proved inefficient, the \texttt{cURL} project stopped paying bounties; submitting security reports is nonetheless possible and no negative impact on the amount of submitted security reports has been observed~\cite{lwn_1}. But the story does not end here. The maintainers started using LLM-based tooling themselves~\cite{lwn_1}. In parallel, the quality of LLM-generated (or aided) security reports has increased to such a level that they are now ``spending multiple hours a day looking at good bug reports''~\cite{stenberg_blog}.

When looking at the overall software ecosystem through the lenses of published CVE numbers\footnote{The Common Vulnerabilities and Exposures reference system for vulnerabilities.}, March and April 2026 reported the highest amount of CVE numbers since the CVE system's inception in 1999. Linux kernel maintainers attributed the rise of their respective CVE numbers to the improved quality of security bug reports, acknowledging the impact of LLMs~\cite{lwn_2}. Mozilla reported in a blog post on using Claude Mythos Preview~\cite{mozilla}, that 423 security fixes were identified in Firefox in April 2026, compared to an average of $21.5$ per month in 2025. Due to \texttt{Claude Mythos}, 271 bugfixes were included in the Firefox 150 release alone.
\vspace{2em} 

\noindent\textbf{The Attacker-Defender Resource Imbalance.}\label{disc:balance}\\

\noindent\textit{The Losers: Bug Bounty Programs and Defenders.}
Our \textit{AI slop case study} illustrates a fundamental asymmetry in resources: LLMs can generate plausible-looking but utterly incorrect security reports in seconds while human experts spend hours or even days reviewing. Consequently, anyone with access to an LLM can flood the security ecosystem with noise, effectively performing a \textit{Denial of Service (DoS)} attack on human expertise. Even worse, attackers could trick over-worked maintainers by secretly injecting new vulnerabilities in maliciously placed patches or bug-fixes~\cite{przymus2026adversarial}. Recent research into this attack vector indicates that the cost of attacking is three orders of magnitude smaller than the costs of defending, and that automated defense systems only have detection rates of around 62\%~\cite{przymus2026adversarial}, further highlighting the resource imbalance and importance of costly human patch-review and oversight (HITL). \\
\vspace{5pt} 

\noindent\textit{The Winners: The Industrialization of Offensive Capabilities.} Research indicates that the main motivation of agentic penetration-testing research is to prepare defenders, with authors often stating that the current efficacy of their prototypes prevents real-world malicious use~\cite{happe2025ethicsusingllmsoffensive}. The latter argument, however, is weakened in the light of recent developments in LLM penetration-testing capabilities (Section~\ref{rise_of_llms}).
The growing imbalance overwhelms the already strained capabilities of defenders, who struggle to patch their systems, detect, and expel attackers from their networks. While defensive AI capabilities will potentially restore the power balance on the long run, agentic AI defenses themselves are vulnerable to attacks, e.g., via prompt injection, further complicating the situation.

\subsection{The Workforce Impact: Deskilling and Labor Displacement}
\label{sec:workforce}

\vspace{5pt} 
\noindent\textbf{Cognitive Erosion and Automation Bias. }In the last section we emphasized the importance of human oversight in patching and defense workflows. The emerge of AI technologies, however, does not only poses a direct, but also an indirect threat to human capabilities: heavy AI use can lead to over-reliance on AI for reasoning, even leading to a loss of critical evaluation capabilities. Studies in other high-stakes fields, such as medicine, have shown that reliance on AI diagnostics can reduce a practitioner's ability to perform independently; other risks are related to reduced cognitive effort and confidence \cite{gerlich2025ai,lee2025impact} as well as reduced skill formation. In cybersecurity, a deskilled operator may lack the expertise to intervene when the AI hallucinates, misinterprets context, or is manipulated by an adversary.

\vspace{5pt} 
\noindent\textbf{The Destruction of the Training Pipeline.} The automation of entry-level tasks potentially removes the traditional training pipelines where new
entrants to the field learn the necessary skills. If AI performs all basic vulnerability scans,
ticket triaging, and log analysis, the industry destroys the training-grounds for creating future senior
experts. We risk creating a generation of script-kiddies who can operate powerful agents
but lack the deep system knowledge required to debug the AI when it fails or to solve novel
problems that the AI cannot handle. Another problem is that operators that are over-relying on LLMs for making their judgment calls; conformity with the AI's judgment could lead to approving potentially destructive operations despite the deployment of HITL as a safeguard~\cite{Liel02102025}.

As of today, the ISC2 estimates a global workforce gap of 4.7 million experts in cybersecurity. Paradoxically, AI may exacerbate this crisis in the long term~\cite{isc2}: as already pointed out in Bainbridges~\cite{bainbridge1983ironies} seminal paper \textit{Ironies of Automation} in 1983: ``\textit{There is some concern that the present generation of automated systems, which are monitored by former manual operators, are riding on their skills, which later generations of operators cannot be expected to have}''
Hence, over-reliance on AI agents for penetration-testing and log analysis, as well as automation-bias --- the tendency to trust an automated suggestion even when it contradicts one's own judgment --- may erode the cybersecurity skills of future professionals.

\subsection{The Regulation Perspective: Governing Pen-testing Agents}\label{sec:regulation}

As the risks of AI become apparent, regulators have begun designing legal frameworks to govern their development and use. The European Union's AI Act is the most comprehensive attempt to regulate AI based on a tiered risk taxonomy in which AI Systems are classified according to the potential harm they pose to health, safety, and fundamental rights. In this taxonomy, \textit{unacceptable risk} use-cases are prohibited outright. This includes systems that manipulate human behavior to cause physical or psychological harm. In an offensive context, this restricts the use of AI agents e.g. for psychological profiling or social engineering in red-teaming. Next, \textit{high risk} use-cases within critical infrastructure, education, employment, and essential public services must comply with strict standards for accuracy, robustness, and cybersecurity throughout their lifecycle. AI systems in high risk use cases must be transparent and incorporate measures for human oversight to prevent automation bias, thereby reinforcing the Human-in-the-Loop requirement. Finally, \textit{General-Purpose AI} models can perform a wide range of tasks, thus carry systemic risks. Such models must be evaluated for and mitigate misuse and adhere to transparency as well as copyright rules. This implies that model-providers must perform cyber-security capability evaluations before releasing their models.

\vspace{5pt} 
\noindent\textbf{Scientific-Research Exemptions in the AI Act.} A critical aspect of the AI Act for the security community is the exemption for research and
development. Article 2(6) states that the regulation does not apply to AI systems or models
\textit{specifically developed and put into service for the sole purpose of scientific research and development}. Additionally, Article 2(8) provides a general exemption for research and testing activities occurring prior to market placement. This research loophole allows academic institutions and security firms to develop offensive prototypes to better prepare defenders. However, the practical application of these exemptions remains uncertain. If an agent is released under an open-source license, it could be considered as being ``put into service'' and nonetheless fall under the scope of the AI Act if classified as ``high-risk''.

Moreover, legal frameworks necessarily lag behind technical capability, yielding a classic Collingridge dilemma~\cite{collingridge1980social}: in the early phase of an emerging technology, accurate technological impact assessments are not possible due to missing historic data. This leads to unregulated practice and deadlocked status-quo, whose ex-post regulation is hard. Ethical research standards balancing legitimate scientific inquiry and the creation of dangerous, unregulated tools, potentially prevent such dilemmas. 

\section{Recommendations}
\label{sec:recommendations}
The preceding analysis reveals that moral responsibility and leverage to shape outcomes is distributed unevenly across stakeholders. This section translates ethical principles into actionable recommendations tailored to each stakeholder's agency. We also acknowledge the uncomfortable reality that black hats operate off ethical norms, making voluntary codes inapplicable. Our recommendations therefore focus on researchers, tool-makers, policymakers, as well as users and defenders. While a subset of our recommendations are already best-practices, we argue, that the potential agent-powered industrialization of offensive security increases the importance and pressure of implementing those practices. Table~\ref{tab:best-practice} summarizes established best-practice recommendations inherited from pre-agentic cybersecurity and AI governance traditions. We focus the remainder of this section on recommendations specific to autonomous agentic penetration-testing systems.

\begin{table}[h!]
\centering
\caption{Established Best-Practice Recommendations That Are Orthogonal to the Use of Autonomous Agentic Systems for Offensive Security.} \label{tab:best-practice}
\begin{tabular}{p{12.3cm}}
\toprule
\textbf{Researchers.}\\
\midrule
\begin{scriptsize}\textbf{Prioritize Defensive Research. }Research prioritizing defender uplift over attacker advantage should be favored, though the distinction is not always clear: strong defenses require understanding attacker behavior. When optimizing both simultaneously, defense must be the explicit focus, with attacker components behind structured access controls.\end{scriptsize} \\
\begin{scriptsize}\textbf{Adopt Responsible Disclosure.} Vulnerabilities identified by autonomous agents must follow standard responsible disclosure principles. The method of discovery (LLM vs. human) does not alter ethical obligations to affected vendors.\end{scriptsize}  \\
\begin{scriptsize}\textbf{Keep Humans in the Loop.} Retain HITL oversight on agent execution outside contained test-beds. Without establishing community norms now, autonomous hacking risks normalizing ``vibe-hacking''.  \end{scriptsize} \\
\begin{scriptsize}\textbf{Disclose Dual-Use Implications}. Publication venues should require structured dual-use assessments alongside methods sections, addressing defender-vs-attacker ratio, safeguards, artifact reversibility, and deployment accessibility. This mirrors emerging DURC norms in life sciences, shifting ethical responsibility to authors at submission.\end{scriptsize}\\
\midrule
\textbf{Defenders \& Maintainers. } \\\midrule\begin{scriptsize} \textbf{Prepare for Increased Vulnerability-Report Volume.} Document disclosure procedures and practice HITL for updates and disclosure. If running a bug-bounty program, investigate non-monetary incentives to filter out LLM-generated low-quality submissions. \end{scriptsize}\\
\begin{scriptsize} \textbf{Minimize Attack Surface and Invest in Good Security Hygiene.} LLM-powered attack industrialization makes this critical: vulnerabilities in undeployed or removed code cannot be exploited. The Linux kernel community exemplifies this by removing unmaintained code ~\cite{phoronix_1}, enabling defenders to focus on deployed components.\end{scriptsize}\\
 \midrule
\textbf{Policy-Makers. }\\\midrule\begin{scriptsize}\textbf{Mandate HITL in Regulated Settings.} Mandatory HITL has limits against illegitimate actors (black-hats), but is enforceable for regulated entities through existing frameworks (NIS2, DORA in the EU; CIRCIA in the US), creating market incentives for vendors.\end{scriptsize} \\
\begin{scriptsize}  \textbf{Support OSS Maintainers and Infrastructure.} The \texttt{cURL} case  illustrates structural under-investment of the OSS ecosystem. The maintainer-economy cannot bear an LLM-amplified increase in security reports without targeted financial and operational support. \end{scriptsize}\\
\bottomrule
\end{tabular}
\end{table}
\subsection{For Researchers, Model-Makers, and Tool-Builders}

Academic researchers and AI tool-makers occupy a consequential position in this ecosystem: Operating at low budgets with strong publication incentives, their artifacts become primitives that other actors integrate.

\vspace{5pt} 
\noindent\textbf{Logging, Audits and Explainability.} 
Incorporate logging tools, facilitating traceability and ex-post incident analysis. Research into explainable AI and human-agent interfaces to enable efficient HITL oversight and tool transparency.

\vspace{5pt} 
\noindent\textbf{LLM-Substrate Disclosure and Evaluation.} Tool authors must disclose the models tested, their concrete versions, and evaluate across at least two models to distinguish scaffold contribution from model contribution.

\vspace{5pt} 
\noindent\textbf{Threat Model your Scaffold/Harness.} Tool authors must publish a safety review covering which actions the scaffold is designed to refuse, which it cannot due to model indeterminateness, and an explicit threat model for downstream scaffold modification. Employ purpose-built threat modeling approaches such as OWASP MAESTRO~\cite{Huang2025}. Academic publishers and venues should mandate inclusion of a threat model and impact analysis for submissions.

\vspace{5pt} 
\noindent\textbf{For Model-Makers.} We advocate for open-weight models for defensive accessibility and transparency reasons. Yet, if you are creating LLMs with offensive cybersecurity capabilities, we recommend keeping them closed-weight and making them available through structured access, implementing a thorough know-your-customer (KYC) validation. Independent of the access mode, implement safeguards and train your model with ethical guidelines. While safeguards can be bypassed, at least they make attacking the system harder. Be transparent on your ethical guidelines and training data.

\subsection{For Society and Policy-Makers}
Regulatory leverage is limited: it cannot constrain black-hat actors, it moves slow, and unilateral national regulation creates jurisdictional arbitrage. Within these constraints, we suggest directions where policy can plausibly help.

\vspace{5pt} 
\noindent\textbf{Build Alternative Training Pipelines for AI-Era Cybersecurity.} The destruction of the training pipeline argument presents a workforce risk that is largely outside the reach of individual employers. If junior tasks are economically destroyed by autonomous agents, society needs deliberate substitute pathways. National cybersecurity strategies should recognize AI-oversight competence as a distinct skill category worth cultivating.

\vspace{5pt} 
\noindent\textbf{Supranational Regulation.} Multilateral frameworks are necessary but must learn from the 2013–2017 Wassenaar cyber-tools episode. Any export-control regime should include explicit research exemptions, narrowly define controlled artifacts to avoid criminalizing vulnerability disclosure, and involve the security community in drafting. Clear rules and boundaries must enable ethical research and dissemination within defensive parties while slowing malicious adversarial uptake of these emerging technologies.

\vspace{5pt} 
\noindent\textbf{Development of Locally-Trained Models}. Public funding and support is needed for creating locally-trained models, esp. for under-represented demographics. This allows for stricter control of training-data composition as well as regional deployment policies, independent from the current US–China duopoly.

\section{Conclusion}
The shift from deterministic, human-operated offensive tools to autonomous LLM-driven penetration-testing agents obfuscates the boundaries of moral agency, creates complex supply-chain dependencies, and challenges existing ethical frameworks in cybersecurity \cite{Formosa2021}. At the same time, we experience the attacker-defender imbalance, and the power-asymmetry between closed-source model-makers and their users. In this tension, we call for a posture of methodological humility. Offensive security has, for thirty years, navigated dual-use dilemmas that regulation could not solve. Therefore, living up to the spirit of \textit{ethical} hacking, it is our community responsibility to drive regulation and governance, as well as to intensify dialogue with the ethics community, governance bodies and affected stakeholders. As we evolve faster under high uncertainty about capability trajectories, and with stakeholders whose interests are increasingly globally asymmetric, we must recommit to human oversight and moral agency.

\begin{credits}
\vspace{5pt} \noindent\textbf{\ackname} LLMs were used for language improvement; outputs were reviewed for accuracy, but not for literature research, methodology, or evaluation. 
\end{credits}

 \bibliographystyle{splncs04}
\bibliography{bibliography}

\end{document}